# An optimized protocol for single cell transcriptional profiling by combinatorial indexing

Beth K. Martin[1], Chengxiang Qiu[1], Eva Nichols[1], Melissa Phung[1,2], Rula Green-Gladden[1,3], Sanjay Srivatsan[1,4], Ronnie Blecher-Gonen[5], Brian J. Beliveau[1], Cole Trapnell[1,7,8], Junyue Cao[6], Jay Shendure[1,7,8,9]

1. Department of Genome Sciences, University of Washington, Seattle, WA, USA.
2. Department of Biology, Case Western Reserve University, Cleveland, OH, USA.
3. Division of Hematology/Oncology, Seattle Children's Hospital, Seattle, WA, USA.
4. Medical Scientist Training Program, University of Washington, Seattle, WA, USA.
5. The Crown Genomics Institute of the Nancy and Stephen Grand Israel National Center for Personalized Medicine, Weizmann Institute of Science, Israel.
6. Laboratory of Single-Cell Genomics and Population Dynamics, The Rockefeller University, New York, NY, USA.
7. Brotman Baty Institute for Precision Medicine, Seattle, WA, USA.
8. Allen Discovery Center for Cell Lineage Tracing, Seattle, WA, USA.
9. Howard Hughes Medical Institute, Seattle, WA, USA.

**Abstract**

Single cell combinatorial indexing RNA sequencing (sci-RNA-seq)[1,2] is a powerful method for recovering gene expression data from an exponentially scalable number of individual cells or nuclei. However, sci-RNA-seq is a complex protocol that has historically exhibited variable performance on different tissues, as well as lower sensitivity than alternative methods. Here we report a simplified, optimized version of the three-level sci-RNA-seq protocol that is faster, higher yield, more robust, and more sensitive, than the original sci-RNA-seq3 protocol, with reagent costs on the order of 1 cent per cell or less. We showcase the optimized protocol via whole organism analysis of an E16.5 mouse embryo, profiling ~380,000 nuclei in a single experiment. Finally, we introduce a "tiny sci-*" protocol for experiments where input is extremely limited.

## Introduction

Single cell combinatorial indexing (sci-*) combines *in situ* molecular indexing and a "split-pool" framework in order to uniquely label an exponentially scalable number of cells or nuclei with a unique combination of nucleic acid barcodes. Following its demonstration in the context of chromatin accessibility[3,4] in 2015, we and others have additionally developed sci-* methods for profiling gene expression[1,2,5–7], genome sequence[8,9], genome architecture[10], genome-wide methylation[11], co-assays of mRNA and chromatin accessibility[12], transcriptional dynamics[13], transcription factor occupancy[14,15], surface proteins[14], small molecule exposures[16] and spatial locations[17], all at single cell resolution. A sci-* method for profiling gene expression, sci-RNA-seq, was first reported in 2017[2], and an improved three-level version, sci-RNA-seq3[1,2], in 2019. Amongst other applications, sci-RNA-seq3 has been applied to generate the largest atlases of single cell gene expression for both mouse[2] (~2 million cells) and human[18] (~4 million cells) to date. Both of these datasets were largely generated within a single lab, each within a few weeks and by a single individual. Nonetheless, the underlying protocol remains cumbersome. Here, we describe the culmination of our extensive efforts to simplify and optimize it.

Briefly, the sci-RNA-seq3 protocol (**Figure 1**) starts by allocating fixed cells or nuclei to the wells of one or more 96-well plates. The first index is introduced during reverse transcription with barcoded oligo-dT primers. Cells or nuclei are then pooled and split to a new set of one or more 96-well plates. The second index is ligated onto the end of the first index, and then the cells or nuclei are pooled and split again. In the third set of plates, second strand synthesis occurs and the double-stranded product is then tagmented with Tn5 transposase. PCR amplification adds the third index and, finally, the library is purified and sequenced.

Here we focus on describing the optimized sci-RNA-seq3 protocol as applied to nuclei, rather than cells. In order to expand the range of tissues that can be processed with sci-RNA-seq3, a first set of changes were directed at better neutralizing endogenous RNases found in older embryonic and adult tissues. A secondary consequence of these changes has been an increase in the number of unique molecular identifiers (UMIs) obtained per nucleus. Specifically, diethyl pyrocarbonate (DEPC) in a hypotonic phosphate buffer is now used to inactivate RNases during the lysis step. A checkpoint step is included to ensure no RNase activity still exists before proceeding with the bulk of the protocol, preventing wasted time and reagents. We found that the expensive SuperaseIn RNase Inhibitor that was used in the original sci-RNA-seq3 protocol was ineffective at inactivating the RNases in older tissues, and the change to DEPC alone has had the biggest impact on sci-RNA-seq3 success. A new buffer, 0.3M SPBSTM, replaces the original nuclei suspension buffer, and enables better nuclei recovery during washes and spins. DSP/methanol fixation replaces the need for a separate permeabilization step and results in more UMIs per cell compared to paraformaldehyde fixation in the original protocol. We have eliminated the USER step, and deoxyU is no longer needed in the ligation primer. These changes allow transcripts to be recovered from RNase-rich tissues that were previously problematic for sci-RNA-seq3, while also mitigating nuclei losses for precious samples, as more nuclei are able to tolerate the entire process.

The optimized sci-RNA-seq3 protocol presented here is scalable and is written here as the basic "1-plate version", with 96 reverse transcription indexes, 96 ligation indexes, and 96 PCR indexes. In our hands, a 96 x 96 x 96 experiment typically nets ~100,000 nuclei. However, there are usually enough nuclei to fill an additional 3 plates for the final round of indexing, such that it is straightforward to boost the number of nuclei profiled to ~400,000 (*i.e.* 96 x 96 x 384). Further scaling to 384 x 384 x 384 facilitates straightforward profiling of 1 million or more cells per experiment. As with all sci-* protocols, cells or nuclei from different samples can be deposited to different wells during the first round of indexing, facilitating multi-sample processing while minimizing batch effects[2].

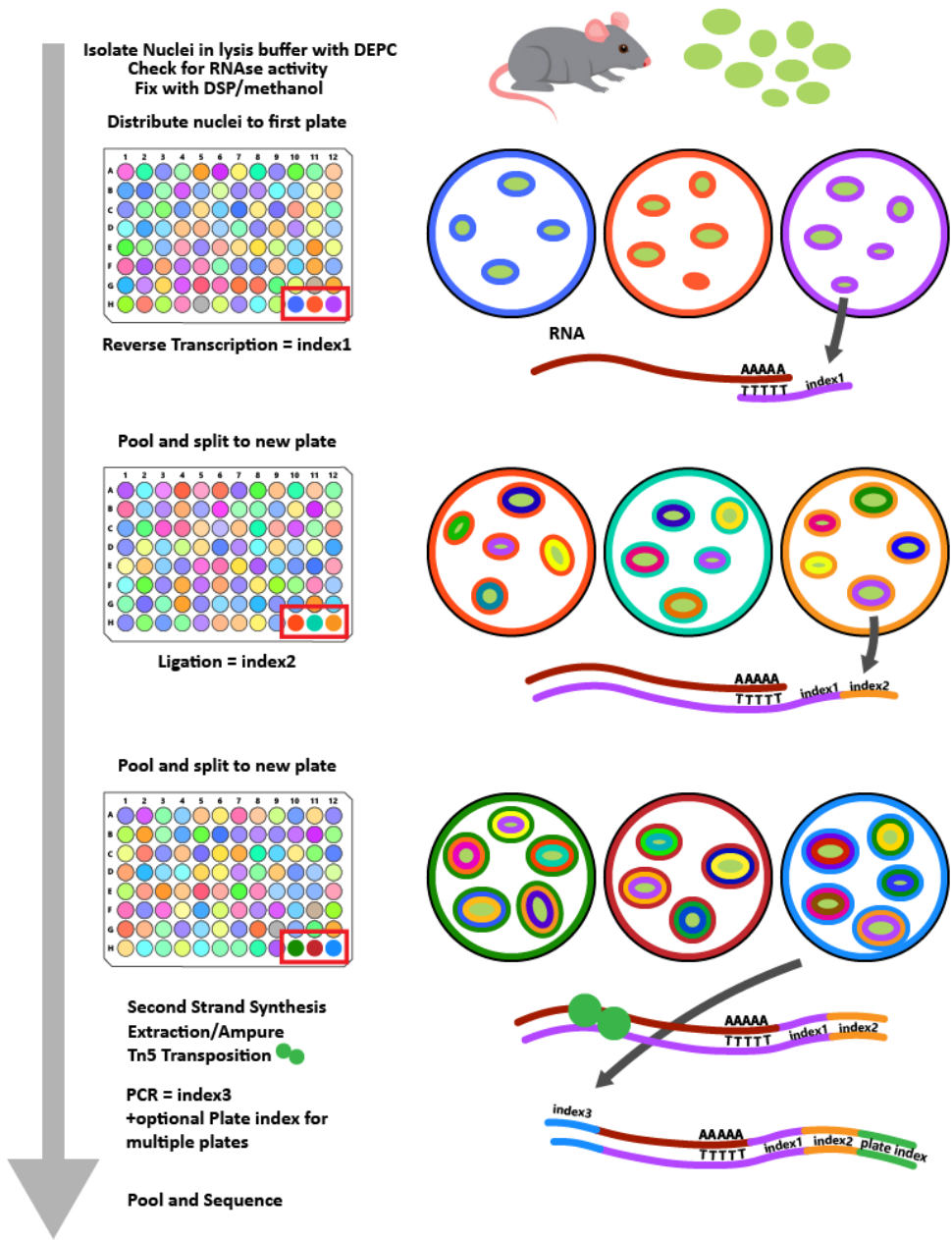

**Figure 1. Summary of optimized sci-RNA-seq3 method.** Nuclei are isolated in lysis buffer with DEPC, then fixed with DSP and methanol. Nuclei are then distributed to a 96-well plate for reverse transcription, where the first index is introduced. If desired, nuclei from different samples can be deposited to different wells during this first round of indexing, facilitating multi-sample processing while minimizing batch effects. After indexed reverse transcription, the nuclei are pooled and split into a new plate to add a second index via ligation, and then the nuclei are pooled and split again. After second-strand synthesis and tagmentation, the third index is added by PCR. Finally, the library is purified and sequenced.

| Enzyme | Cost/µl | Original amount/plate | Original cost/plate | Optimized amount/plate | Optimized cost/plate |
|---|---|---|---|---|---|
| Superscript IV | $7.11 | 205 µl | $1458 | 55 µl | $391 |
| RNaseOUT | $1.56 | 205 µl | $320 | not used | $0 |
| SuperaseIN | $1.00 | ~250 µl | $250 | not used | $0 |
| Quick Ligase | $2.59 | 215 µl | $557 | not used | $0 |
| T4 DNA Ligase | $1.04 | not used | $0 | 65 µl | $68 |
| USER | $1.21 | 110 µl | $133 | not used | $0 |
| Second Strand Synthesis | $2.95 | 73.3 µl | $216 | 35 µl | $103 |
| AmpureXP | $0.014 | 3840 µl | $53 | not used | $0 |
| Tn5 | $24.50 | 0.92 µl | $22 | 4.6 µl | $112 |
| NEBNext 2X PCR mix | $0.058 | 1920 µl | $111 | 2200 µl | $127 |
| **Total enzyme cost per plate (96x96x96)** | | | $3,120 | | $801 |
| **Total enzyme cost per 4 plates (384x384x384)** | | | $12,480 | | $3,204 |

**Table 1. Cost comparison between original[1] vs. optimized sci-RNA-seq3 protocols.** We focus on the most expensive reagents here (all enzymes) as other reagent costs are comparatively negligible.

**Materials**

**Reagents**

Dithiobis (succinimidyl propionate) (DSP, Lomant's Reagent; Thermo Fisher, cat. no. 22586)

Methanol (Millipore Sigma, cat. no. 494437-2L)

DMSO (Millipore Sigma, cat. no. D2438-5X10ML)

Sodium Phosphate Dibasic (Millipore Sigma, cat. no. S3264-250G)

Sodium Phosphate Monobasic Monohydrate (Millipore Sigma, cat. no. 71507-250G)

Potassium Phosphate Monobasic (Millipore Sigma, cat. no. P9791-100G)

Sodium Chloride (Millipore Sigma, cat. no. S3014-500G)

Potassium Chloride (Millipore Sigma, cat. no. P9541-500G)

Magnesium Chloride solution 2M (Millipore Sigma, cat. no. 68475-100ML-F)

Igepal CA-630 (Millipore Sigma, cat. no. I8896-50ML)

Bovine Serum Albumin 20mg/ml (New England Biolabs, cat. no. B9000S)

DEPC (Diethyl Pyrocarbonate) (Millipore Sigma, cat. no. D5758-25ML) **CAUTION** handle DEPC, and samples containing it, in a fume hood

Sucrose (VWR, cat. no. 97061-428)

TritonX-100 (Millipore Sigma, cat. no. T8787-100ML)

Tween 20 (Thermo Fisher BP-337-100)

10X Dulbecco's Phosphate Buffered Saline (10XDPBS; Thermo Fisher, cat. no. 14200075)

Superscript IV Reverse Transcriptase (Thermo Fisher, cat. no. 18090200)

T4 DNA Ligase (New England Biolabs, cat. no. M0202L)

Tagmentase (Tn5 transposase) - unloaded (Diagenode Cat# C01070010-20)

Tn5-N7 oligo (5′-GTCTCGTGGGCTCGGAGATGTGTATAAGAGACAG-3′, Eurofins, High-Purity Salt-Free)

Mosaic End (ME) oligo (5′-/5Phos/CTGTCTCTTATACACATCT-3′, Eurofins, High-Purity Salt-Free)

NEBNext® mRNA Second Strand Synthesis Module (New England Biolabs, cat. no. E6111L)

NEBNext high fidelity 2x PCR master mix (New England Biolabs, cat. no. M0541L)

dNTP mix (New England Biolabs, cat. no. N0447L)

Agencourt AMPure XP (Beckman Coulter, cat. no. A63882)

YoYo dye (Thermo Fisher, cat. no. Y3601)

RNaseAlert kit (IDT, cat. no. 11-02-01-02)

RNaseZap (Thermo Fisher, cat. no. AM9780)

Elution buffer (EB, 10mM Tris pH8.5, Qiagen 19086)

Protease (Qiagen, cat. no. 19157)

sci-RNA-seq3 indexed primer plates at 10µM dilution (standard desalting for purification, random bases do NOT need hand-mixing):

Plate(s) of indexed oligo-dT primers (5′-/5Phos/CAGAGCNNNNNNNN[10bpRTindex]TTTTTTTTTTTTTTTTTTTTTTTTTTTTTT-3′, where "N" is any base; IDT)

Plate(s) of indexed ligation primers (100µM, 5'- GCTCTG[9bp or 10bp barcode A]TACGACGCTCTTCCGATCT[reverse complement of barcode A]-3')

PCR P5 primers (this primer doesn't need to be indexed if you only do one plate of pcr) (5′-AATGATACGGCGACCACCGAGATCTACAC[i5]ACACTCTTTCCCTACACGACGCTCTTCCGATCT-3′, IDT)

Plate of Indexed PCR P7 primers (5′-CAAGCAGAAGACGGCATACGAGAT[i7]GTCTCGTGGGCTCGG-3′, IDT)

Qubit dsDNA HS quantitation kit (Thermo Q32851)

## **Equipment**

Hammer

Low-bind tubes (Eppendorf, cat. no. 22431021)

Refrigerated centrifuges that hold 1.5 ml microfuge tubes, microwell plates, 15 ml and 50 ml conical tubes

Chemical fume hood

Multichannel pipettes and tips

FloMi filter 40µM (VWR, cat. no. 10032-802)

Falcon cell strainer 40µM (VWR, cat. no. 21008-949)

Pestle for cell strainer (Midsci, cat. no. SG-PEST)

96-well plates (Eppendorf, cat. no. 951020401)

96-well LoBind plates (Eppendorf, cat. no. 30129512)

Thermomixer

Sonicator (Diagenode Bioruptor)

Cell counter with GFP channel, or a hemocytometer that allows visualization with GFP

### **Reagent Setup**

Prepare necessary buffers and fixatives. For lysis, a hypotonic version of PBS is used, with MgCl2 for nuclei stability, igepal for the detergent, and DEPC to neutralize any RNases that are released. Nuclei are washed and pooled in a PBS-based buffer with sucrose for osmolarity and TritonX to help with pelleting during centrifugation. Nuclei are fixed with a combination of DSP and methanol.

10X-PBS-hypotonic stock solution

Mix 5.45g Na2HPO4 (dibasic), 3.1g NaH2PO4-H20, 1.2g KH2PO4, 1g KCl, 3g NaCl in nuclease-free water and bring to a final volume of 500 ml. This stock solution will be ~pH6.8, but when diluted to 1X should end up at pH7.0-7.4. The buffer can be stored at room temperature.

Lysis buffers

There are two lysis buffers to choose from, depending on the tissues involved. Older mouse embryos perform better with buffer A, which is sucrose-based and lacks BSA. Younger embryos and isolated tissues will clump less with the BSA-based buffer B.

Hypotonic Lysis buffer solution A - used for mouse embryos 13.5 and older

Mix 5ml of the 10X-PBS-hypotonic stock solution, 5.7g sucrose, 75 µl of 2M MgCl2, and nuclease-free water to a final volume of 50 ml to make the lysis base solution. *Right before lysis*, for every 1 ml of lysis buffer needed, add 2.5 µl 10% igepal and 10 µl DEPC, then vortex solution to disperse the DEPC throughout. Example: If a sample needs 5 ml of lysis buffer, take a 5 ml aliquot of lysis buffer stock solution, add 12.5 µl 10% igepal, 50 µl DEPC. Keep buffer on ice.

**CAUTION** DEPC needs to be used in a fume hood.

**CRITICAL** DEPC has a short half life in aqueous solutions, so it's important to add it to the buffer just before the cells are added.

Hypotonic Lysis buffer solution B - used for Tiny Sci, younger mouse embryos.

Mix 5ml of the 10X-PBS-hypotonic stock solution, 75 µl of 2M MgCl2, and nuclease-free water to a final volume of 50 ml to make the lysis base solution. *Right before lysis*, for every 1 ml of lysis buffer needed, add **40ul BSA** (20mg/ml), 2.5 µl 10% igepal and 10 µl DEPC, then vortex solution to disperse the DEPC throughout. Example: If a sample needs 5 ml of lysis buffer, take a 5 ml aliquot of lysis buffer stock solution, add 200ul BSA, 12.5 µl 10% igepal, 50 µl DEPC. Keep buffer on ice.

**CAUTION** DEPC needs to be used in a fume hood.

**CRITICAL** DEPC has a short half life in aqueous solutions, so it's important to add it to the buffer just before the cells are added.

0.3M SPBSTM (Sucrose PBS TritonX MgCl$_2$)

This is the main buffer used throughout the protocol for washing and diluting nuclei. Dissolve 28.5g sucrose in 25ml 10X PBS (regular PBS, not the hypotonic version) and 125ml nuclease-free water (about half the volume of water you'll need). Once the sucrose has dissolved, add 2.5ml 10% TritonX-100, 375 µl of 2M MgCl$_2$, and more water to the final volume of 250ml. Store this buffer at 4°C.

DSP 50 mg/ml stock

Dissolve a 50 mg vial of DSP in 1 ml of anhydrous DMSO (use a new vial of DMSO), as DSP will precipitate in aqueous solutions.

Dispense into 100 µl aliquots and store at -80°C.

Yoyo-1 dye for counting

Dilute 1 µl of Yoyo-1 dye in 1 ml of 0.3M SPBSTM in a dark or amber microfuge tube, and store the reagent at 4°C. This will be used to dilute nuclei for counting.

Annealing N7 oligos

Tn5-N7           5′-GTCTCGTGGGCTCGGAGATGTGTATAAGAGACAG-3′

Mosaic End (ME)  5′-[phos]CTGTCTCTTATACACATCT-3′

Resuspend both oligos to 100µM in annealing buffer (50 mM NaCl, 40 mM Tris-HCl pH8.0).

Mix one volume of Tn5-N7 with one volume of ME. This creates a working stock at 50 µM. Anneal them with the following PCR program: 95°C 5min, cool to 65°C (0.1°C/sec), 65°C 5min, cool to 4°C (0.1°C/sec). Store annealed oligos at 4°C or aliquot and freeze at -20°C.

N7-loaded Tn5

Tagmentase (Tn5 transposase) - unloaded (Diagenode Cat# C01070010-20).

To 20 µl of Tn5, add 20 µl of annealed N7 oligos. Place in a thermomixer and shake at 350 rpm, 23°C, 30 minutes.

Add 20 µl of glycerol. Store at -20°C.

Tagment DNA (TD) buffer (2X)

To 38.75 ml of nuclease-free water, add 1 ml 1M Tris pH7.6, 250 µl of 2M MgCl2, 10 ml dimethylformamide. Final volume is 50 ml. Make 550 µl aliquots and store at -20°C.

Indexed Primer Plates

Primers for reverse transcription, ligation, and PCR indexing steps are at 100 µM. Working dilutions are made to 10 µM in EB, and kept at 4°C.

### 10% Igepal

Dilute 5 ml Igepal in 45 ml nuclease-free water. Store at room temperature.

### 10% TritonX-100

Dilute 5 ml TritonX-100 in 45 ml nuclease-free water. Store at room temperature.

### 10% Tween20

Dilute 5 ml Tween20 in 45 ml nuclease-free water. Store at room temperature.

### Protease

Add 7ml water to a bottle of lyophilized Qiagen protease (Qiagen #19157). Make 200ul aliquots and store -20C up to 6months. Don't freeze/thaw.

**Protocol**

**General Notes**

This protocol is written for a small experiment with one plate of primers at each indexing round (which typically yields ~100,000 single cell profiles). It can be scaled up easily to use as many as 4 plates of primers at each round (which would be expected to yield over ~1 million single cell profiles) - just multiply each step by 4, combine all 4 plates when pooling, and put more nuclei (4000/well) into the final plates.

**Everything is to be kept cold at all times**. Have lots of ice ready, pre-cool centrifuges to 4°C. Pre-cool all tubes on ice before you put nuclei in them.

Clean workspace and fume hood area ahead of time - wipe pipettes, racks, centrifuges down with RNAseZap, change gloves often.

Spin the primer plates down before opening, but don't spin plates with nuclei until the second strand synthesis stage.

Use low-bind microcentrifuge tubes for the nuclei.

**Timeline**

There are three main sections to the protocol that can be split over multiple days.

1) **Day 1** - Nuclei preparation and fixing. Fixed samples can be stored at -80°C.

2) **Day 2** - Reverse Transcription, Ligation, Second Strand Synthesis (can set this to go overnight if you are picking things back up the next day, otherwise do this step on third day)

3) **Day 3** - Extraction, Tagmentation, PCR, final library cleanup and load on sequencer.

**Counting Nuclei**

Nuclei prepared from frozen tissue are not simple to count, as debris can interfere with interpretation. Staining with yoyo-1 dye helps to discern nuclei from debris and is visualized on the GFP channel. Mix 10 μl of diluted yoyo-1 (see reagent setup) with 10 μl of nuclei (2 fold dilution) or do a 10-20-fold dilution if the nuclei are concentrated.

For the example experiment described below, nuclei were counted *manually* on a Countess Cell counter, as the automatic cell counting was unreliable for nuclei. On the GFP channel, with the view zoomed all the way out, yoyo-1-stained nuclei were counted in a 6cm x 6cm square. This count was multiplied by the dilution factor x10,000 to get an approximate number of nuclei/ml.

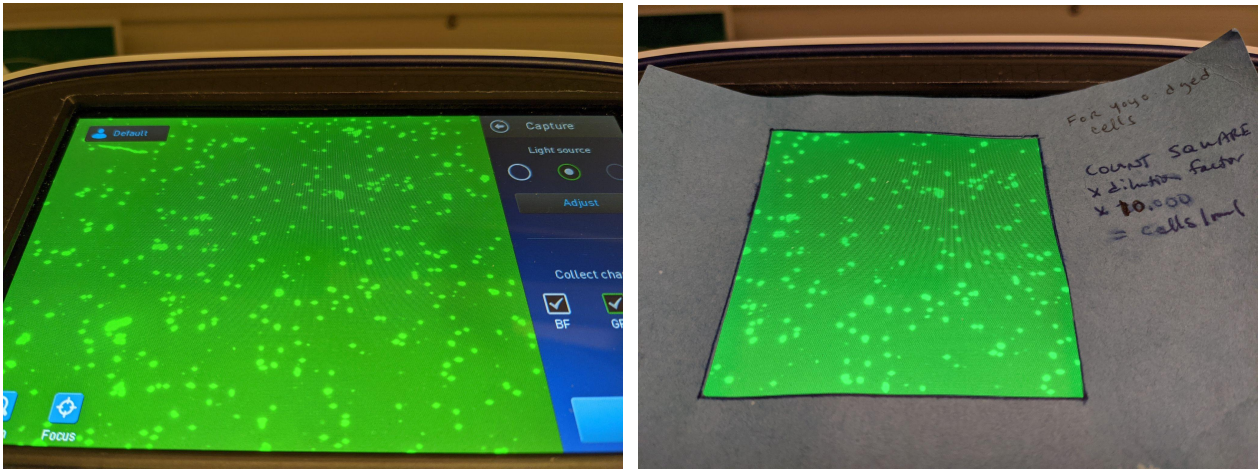

**Figure 2. Yoyo-1 stained nuclei from an E16.5 mouse embryo visualized on a Countess Cell Counter.** Nuclei are counted by hand in a 6cm x 6cm square. The method is a bit inelegant, but in our hands fast and remarkably consistent.

---

[Nuclei Isolation](#)

**Timing: 2 hours**

Note: If you are working with extremely small samples a few millimeters in size (such as E8 mouse embryos), see the "Tiny Sci" section at the end of this protocol for alternative nuclei isolation and fixing steps.

Tissue is dissociated by simply smashing it on dry ice. At this point, RNA is going to be especially vulnerable to RNases that are released by the cells, so a sufficient volume of DEPC-containing lysis buffer is necessary to inactivate them. This is the step that will make or break your experiment, so be sure to do an RNaseAlert check before proceeding to fixation.

1. Remember everything is cold all the time. Prepare two ice buckets with wet ice, a bucket with crushed dry ice to hold your frozen tissues, and a thick, flat slab of dry ice for smashing tissues. Precool a centrifuge that will hold 50ml tubes, and a microfuge, to 4°C.

2. Determine how much lysis buffer you will need for the tissue you will be processing. A E13.5 mouse embryo (~200 mg) works with 5 ml of lysis buffer. A E16.5 embryo (~500 mg) will need 10 ml. An adult mouse heart needs 5 ml. Adult mouse kidneys need 5 ml per kidney. Adult mouse liver needs 20 ml. Adult mouse pancreas needs 15ml. Adult tissues and tissues high in RNases will necessitate a bigger lysis volume. The buffer is inexpensive to make so don't worry about using too much.

3. For every 1 ml of lysis buffer needed, add 2.5 µl 10% igepal, 10 µl DEPC to the hypotonic lysis buffer solution (and 40ul BSA if using lysis buffer B), then vortex solution to disperse the DEPC throughout. **CAUTION** The following steps should be performed in the chemical hood from this point until the DEPC is washed from the sample (step 17). Have complete lysis buffer in a 50 ml tube for each sample on ice ready to go.

4. Fold a piece of aluminum foil so that you have a small pouch with 4 layers of foil on each side. Place this on a slab of dry ice to chill.

5. Place your frozen tissue inside this foil and hold it firmly closed on the dry ice and smash it with a hammer. You want to be gentle enough not to tear the foil, but thorough enough to make a powder of the tissue. Do not let the tissue thaw.

6. Use the foil to guide your powdered tissue into the tube of the lysis buffer. It will stick a bit, pipet some of the lysis buffer from the tube to rinse the sample from the foil into the tube. Try to make sure that the sample is only thawing if it is in lysis buffer.

7. Cap the 50 ml tube and shake to disperse the chunks in the buffer. Let sit on ice for 10 min. Triturate the chunks with a 1 ml pipette tip to help tease them apart a bit.

8. Set up another 50 ml tube on ice with a 40 µm cell strainer on top. Pour your lysate through that – there will still be a lot of chunks. Use a disposable pestle to coax the tissue through the filter. Don't worry about getting all of it through.

9. Take 45 µl sample of the filtered lysate and check for Rnase activity with the IDT RnaseAlert kit. The RNaseAlert will guide you on whether to proceed or not. There should not be any RNase detected, and if there is, you will have to restart with a new sample and adjust either the sample size or the volume of lysis buffer, so that there is enough DEPC to inactivate the RNases. You cannot continue with a sample that has RNases detected at this point, the damage is already done. **TROUBLESHOOTING**

10. While the RNaseAlert sample is incubating, spin down the remainder of the lysate (500xg, 3 min, 4°C). Resuspend the nuclei in 1 ml 0.3M SPBSTM with 10ul DEPC added (or more buffer if there are a lot of nuclei - roughly 1ml buffer per 200mg-500mg starting material at minimum). Keep the nuclei in the 50ml tube. **TROUBLESHOOTING**

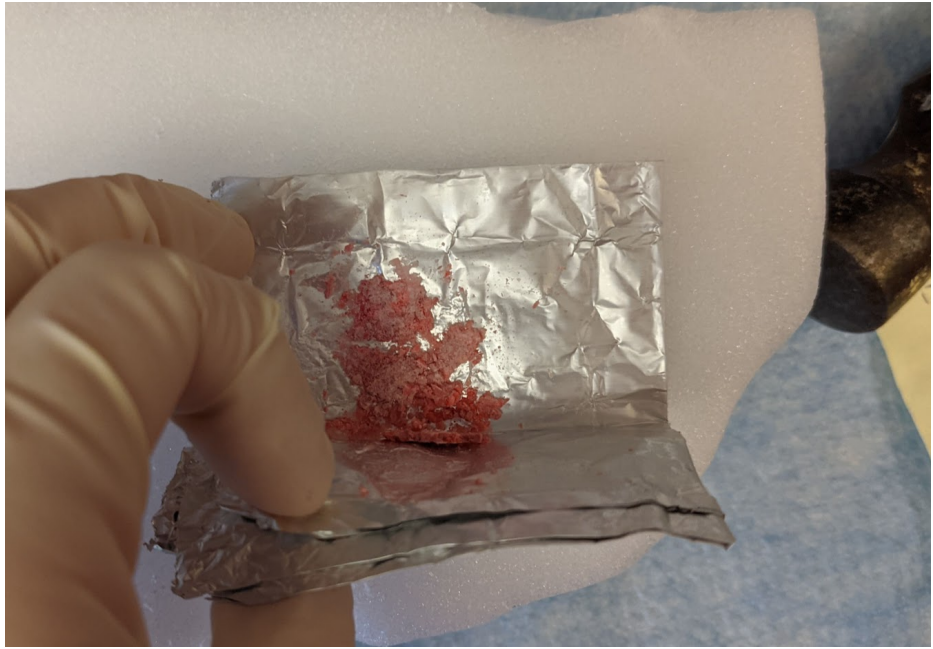

**Figure 3. Smashing tissue in a foil packet on a slab of dry ice with a hammer**. Foil must stay on the dry ice until the powdered tissue is added to lysis buffer.

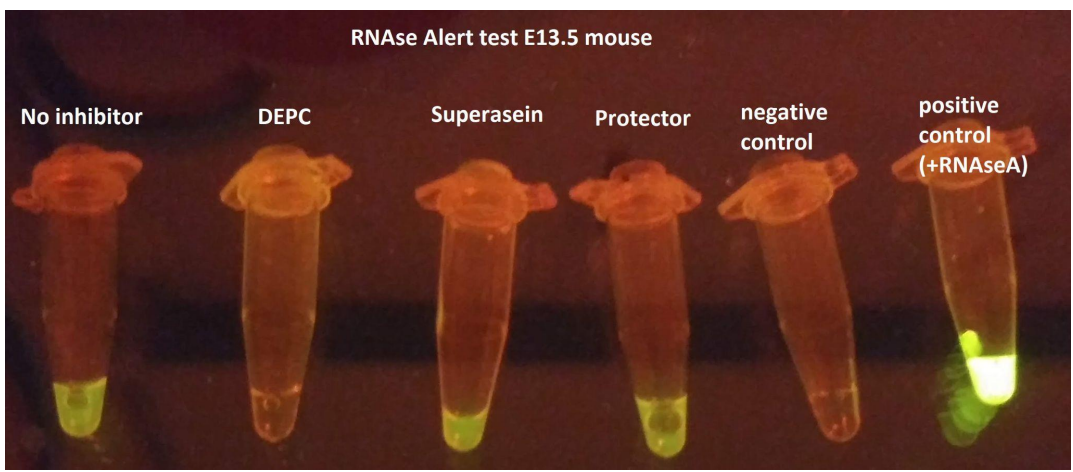

**Figure 4. IDT RNase Alert test of lysed E13.5 mouse nuclei with various RNase inhibitors:** DEPC, Superasin (ThermoFisher), Protector (Roche) (10 µl inhibitor to 1ml of lysis buffer). DEPC added to lysis buffer is the only one that was able to inactivate the RNases.

## Nuclei Fixation

**Timing: 1 hour**

Nuclei are fixed with a combination of DSP and methanol. DSP is an amine-crosslinker that doesn't alter RNA, but unfortunately precipitates in aqueous solutions. Methanol both fixes and permeablizes nuclei well for access to the transcripts, but often leaves nuclei too fragile to tolerate the complete protocol. The two fixatives work well together – the DSP is easily dissolved in the methanol and confers integrity to the fixed cells. The resulting nuclei are stable, accessible, and less prone to clumping compared with the paraformaldehyde fixation in the original protocol.

11. For each sample, prepare fixative: add 100 µl of 50 mg/ml DSP stock solution to 4 ml of ice cold methanol for every 1 ml of nuclei that you are starting with.
12. Add the fixative to the nuclei gradually and swirl to mix.
13. Fix on ice for 15 min, swirling occasionally.
14. Add 2 volumes 0.3 SPBSTM gradually, swirling every few mls, to rehydrate the nuclei.
15. Spin down the nuclei at 500xg 3min 4°C.
16. Carefully remove supernatant and dispose properly. The nuclei pellet is at the bottom and should look a little white-ish from the DSP.
17. Resuspend the nuclei in 1 ml (or more) 0.3M SPBSTM. Triturate gently with a pipette tip to separate nuclei.
18. **OPTIONAL:** If there are obvious clumps at this point that won't tease apart, you will need to sonicate them to break them up. Sonicate on low for 12s only. Spin and resuspend the nuclei in 1 ml 0.3M SPBSTM. **TROUBLESHOOTING**
19. Divide fixed nuclei into aliquots in microfuge tubes. Spin 500xg 3 min 4°C and remove supernatant. Snap freeze tubes in LN2 and store at -80°C.

**Can stop at this point.**

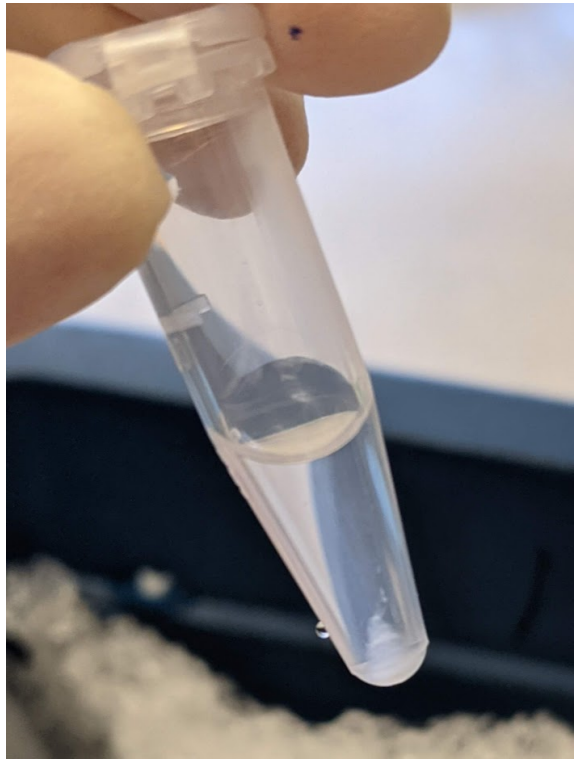

**Figure 5. Nuclear pellet size**. This is approximately the size of the nuclei pellet (~2 million nuclei) needed for 1 plate of reverse transcription. Extra fixed nuclei can be aliquoted and snap-frozen.

## Reverse Transcription (RT)

**Timing: 2-3 hours**

This step will take you longer than you think because you will need to get the right amount of nuclei loaded into the first plate, and especially if you are including multiple samples. Allow for a lot of time and don't rush. For a single sample, you will need 2M nuclei to fill out a plate.

20. Follow the chart below to determine how many starting nuclei you need and their volume.

| cell number: | 2M | 1M | 800K | 500K | 400K | 200K |
|---|---|---|---|---|---|---|
| number of columns: | 12 | 6 | 4 | 3 | 2 | 1 |
| nuclei volume | 500 µl | 250 µl | 170 µl | 125 µl | 85 µl | 42.5 µl |
| 10mM dNTP | 56 µl | 28 µl | 19 µl | 14 µl | 9.5 µl | 4.75 µl |

21. Resuspend an aliquot of frozen nuclei in 500 µl of 0.3M SPBSTM to start. Count. Dilute cells if necessary to get an accurate count. If nuclei are clumpy, even after sonicating, and can't be teased apart with pipetting, then put them over a Flow-mi pipette tip filter before counting. Flow-mi filter is a last resort as it results in nuclei loss, but is helpful if you have an excess of nuclei.
22. Pull out the desired amount of nuclei into a new tube and spin. Remove supernatant and resuspend nuclei in the necessary volume determined by the chart and add the appropriate amount of dNTPs.
23. Aliquot 5 µl nuclei+dNTP mix to each well of the plate on ice.
24. Quickly spin the plate of 3-level RT primers (10µM).
25. Add 2 µl of primer to each well. Don't pipet up and down to mix, just stir gently with the pipet tips.
26. Incubate plate at 55°C for 5 min (heated lid set to 65°C) and then immediately place on ice.

27. While this is incubating, make the reaction mix. Note: we are not including DTT in this mix, as it will undo the DSP crosslinks (it is not necessary for the RT to work).

| RT mix per plate: | each | X120 |
|---|---|---|
| 5X Superscript IV buffer | 2 µl | 240 µl |
| Superscript IV (200 u/µl) | 0.5 µl | 60 µl |
| water | 0.5 µl | 60 µl |

28. Put 3ul of reaction mix into each well (45 µl mix x8 in strip tube for multichannel), stirring gently with tips. 10 µl total now.
29. Incubate 55°C 10 min (heated lid at 65°C) and then immediately place on ice.
30. Ice plates until they are cold (10-15min). Add 5 µl cold 0.3M SPBSTM per well. To maximize recovery, pool wells by using a 12-multichannel with 200ul tips to pipet gently up and down (the pipetting up and down is important to dislodge the nuclei, but try to avoid creating excessive bubbles), and combine each row of the plate into the bottom row. You can use the same tips for the whole plate. Then collect these wells into 2 cold microfuge tubes. (It will be bubbly so it's difficult to squeeze into 1 tube).
31. Spin 500xg, 3min, 4°C. Pellet will be small but you should be able to see it. Remove supernatant.
32. Combine tubes and wash once more in 1 ml cold 0.3 SPBSTM. Spin 500xg, 3min, 4°C. Remove supernatant.

## [Ligation](#)

**Timing: 1 hour**

33. Resuspend nuclei in 1200 µl 0.3M SPBSTM.
34. Distribute 11 µl to each well of a new plate on ice.
35. Quick spin the plate of 3-level ligation primers (10 µM)
36. Add 2 µl of primer to each well. Don't pipet up and down.
37. Make a 3:1 mix of 10X T4 ligation buffer and T4 DNA ligase. (195 µl 10X buffer + 65 µlT4 DNA Ligase)
38. Add 2 µl ligase mix to each well. (32 µl x8 in strip tube for multichannel).15 µl total now.
39. Incubate 20 min at room temperature.
40. Ice plates until cold.

41. Pool wells by using a 12-multichannel to pipet gently up and down (the pipetting up and down is important to dislodge the nuclei), and combine each row of the plate into the bottom row. Then collect these wells into 2 cold microfuge tubes.
42. Spin 500xg, 3min, 4°C. Remove supernatant.
43. Combine the two tubes and wash <u>twice</u> more with 1 ml 0.3M SPBSTM per wash.
44. Resuspend in 1 ml 0.3M SPBSTM to count. If they are clumpy and can't be teased apart with gentle pipetting, put through a tip filter and recount. **TROUBLESHOOTING**

## [Final Distribution](#)

**Timing: 30min**

In the final plate you will want 1000 nuclei/well (or 4000/well if you've scaled up the experiment to 384x384x384). You should have enough nuclei to freeze multiple plates if you like.

45. Make 400 µl 1X Second Strand Synthesis buffer for each plate in the final distribution: Dilute 40 µl 10X Second Strand buffer in 360 µl water to get 1X concentration.
46. Spin down 100K nuclei for each plate desired for the final distribution. (400K per plate if this is a 384x384x384 experiment). For each plate/100K, resuspend in 400 µl 1X Second Strand Synthesis buffer.
47. Put 4 µl nuclei into each well of a regular, not lo-bind, plate on ice.
48. Cover with foil seals and freeze plates at -80°C or proceed with second strand synthesis.

**Can stop here.**

## [Second Strand Synthesis](#)

**Timing: 3 hours**

49. Thaw plate on ice.
50. Make second strand synthesis mix on ice as follows:

| reaction mix per plate: | each | X140 |
|---|---|---|
| water | 0.675 µl | 94.5 µl |
| second strand buffer (10X) | 0.075 µl | 10.5 µl |
| second strand enzyme (20X) | 0.25 µl | 35 µl |

51. Put 1 µl of second strand synthesis mix into each well (17 µl mix x8 in strip tube for multichannel). 5 µl total now.
52. Incubate 16°C 2.5hours. (No heated lid)

**Can stop here, keep the plate at 4°C.**

[Protease Digestion](#)

**Timing: 2 hours**

53. Add 1 µl protease to each well. NOTE: This is NOT proteinaseK. It's important to use Qiagen protease (#19157) because it can be heat-inactivated.
54. Incubate 37C for 30min. Check 1ul on microscope: mix 1ul sample with 2ul of diluted yo-yo1 dye and put this on a slide and check on the GFP channel. You should see whisps of DNA instead of intact nuclei.
55. Heat-inactivate the protease 75C 20min (85C heated lid).

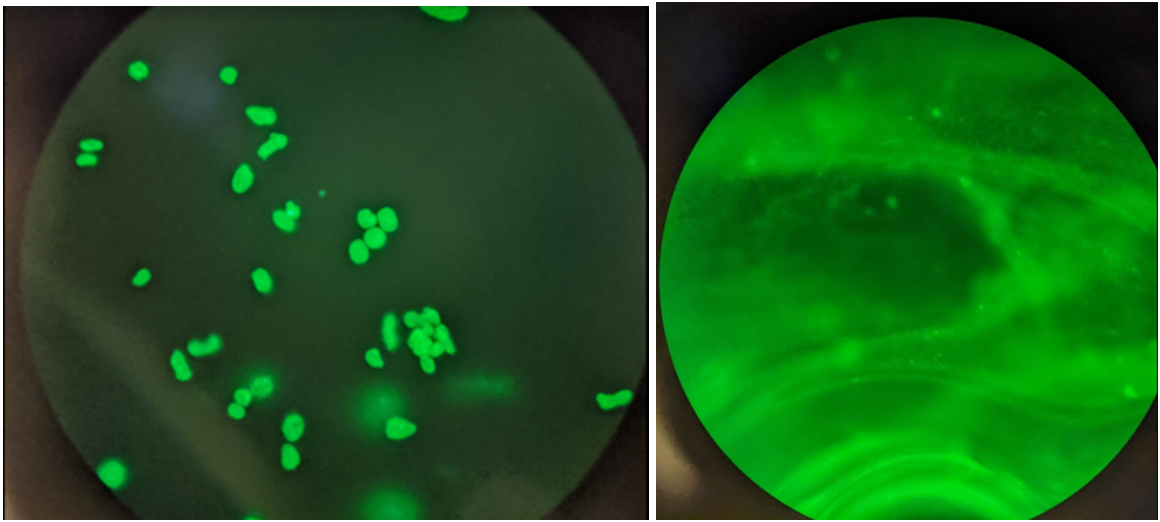

**Figure 6. Visualizing protease digestion of the nuclei.** A. Nuclei after about 10min of protease digestion, swelling and starting to lose integrity. B. Nuclei after 30min of protease digestion; DNA has been released and now the protease can be heat-inactivated.

## Tagmentation

**Timing: 30min**

Tagmentation is done with Tn5 transposase loaded with only the N7 side of the usual Nextera enzyme (see Reagent Setup).

56. Make tagmentation mix as follows:

    | reaction mix per plate: | each | X110 |
    |---|---|---|
    | TD buffer | 5 µl | 550 µl |
    | N7-loaded Tn5 | 0.125 µl | 13.75 µl |

57. Add 5 µl tagmentation mix to each well. ~10-11 µl total now.
58. Incubate 55°C 5min, do not put on ice.
59. Remove the transposases with this buffer (keep at room temperature):

    | reaction mix per plate: | each | X120 |
    |---|---|---|
    | 1% SDS | 0.4 µl | 48 µl |
    | BSA | 0.4 µl | 48 µl |
    | water | 1.8 µl | 216 µl |
    |  | 2.6 µl |  |

60. Add 2.6 µl to each well and mix (39 µl x8 into a strip tube for multichannel).
61. Incubate 55°C 15min.
62. **Quench SDS by adding 2 µl 10% Tween 20 to each well** (bolded this because it is a very easy step to forget)

## PCR amplification

**Timing: 1 hour for PCR, 2 hours for gel purification**

PCR is done with 96 indexed P7 primers. Alternatively, you can also add an indexed primer on the P5 end for multiplexing multiple plates, but this second index is not necessary for single plates.

63. Assemble the PCR master mix:

| reaction mix per plate: | each | X110 |
|---|---|---|
| 2X NEBNext | 20 µl | 2200 µl |
| TruSeqP5-noindex primer (100µM) | 0.2 µl | 22 µl |
| water | 3.2 µl | 352 µl |
| total | 23.4 µl | 2574 µl |

64. Add 2 µl of indexed P7 primers (10µM) to each well.
65. Add 23.4 µl of PCR master mix to each well.
66. Amplify 16 cycles with a pre-extension step in the following program:

| | | |
|---|---|---|
| 1 | 70°C | 3 min |
| 2 | 98°C | 30s |
| 3 | 98°C | 10s |
| 4 | 63°C | 30s |
| 5 | 72°C | 1 min |
| 6 | go to step 3, 15 more times | |
| 7 | 72°C | 5 min |

67. Run 1.5 µl of a few wells on a 6% PAGE gel to check. You should see a smear of products with primer-dimers underneath. We will be isolating a section of the smear, centered on 400bp. **TROUBLESHOOTING**

68. Concentration of library and agarose gel purification: Pool 3 µl of each well and do a 0.8X ampureXP cleanup (230 µl beads). (Save the remaining plate in case you need to redo cleanup or if you anticipate needing more library for a large NovaSeq run). Wash the ampure bead pellet twice gently with 70% EtOH and elute pool in 50 µl. Load this into a single 1 cm well on a 1% agarose gel. Cut out the smear between about 250-600 bp and use the NEB gel extraction kit, using extra dissolving buffer since it will be bigger than a normal slice and run it all through the same purification column. Wash twice with 200 µl NEB wash buffer, elute in 20 µl EB. Quantitate library with Qubit dsDNA HS.

69. Run the library on NextSeq (or NovaSeq depending on final cell numbers or sequencing depth desired), using standard primers. Read1 34 cycles, Index 10 cycles, Read2 48 cycles. If you've also used a P5 index for pcr, then add a second Index read of 10bp.

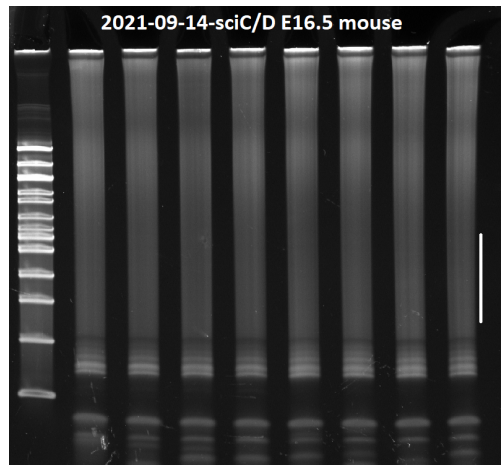

**Figure 7. Post-PCR gel prior to size selection**. 6% non-denaturing PAGE gel with 1.5 µl from a sampling of wells after the PCR, before cleanup. White bar indicates the region that you will size-select for sequencing. First lane: NEB 100bp ladder.

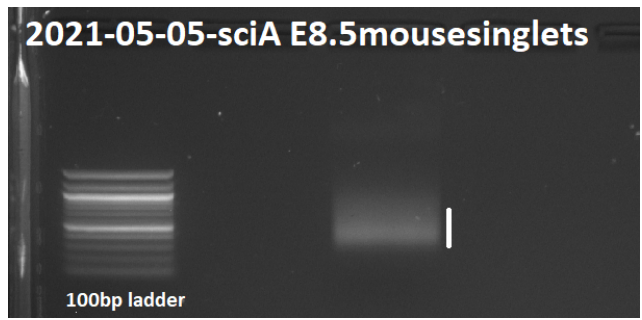

**Figure 8. Size selection**. 1% Agarose gel for size-selecting the library. 3 µl of each well was pooled, ampured 0.8X and eluted in 50 µl to load into a single well on the gel. The bar indicates where to cut out the smear of products. The ampure has taken care of removing the primer-dimers seen on the previous gel, so gel size-selection is optional, but gives a tighter size distribution for quantification of the library and sequencing. First lane: NEB 100bp ladder.

**Troubleshooting Guide**

| Step | Problem | Possible Reason | Solution |
|------|---------|-----------------|----------|
| 9 | RNase check is positive for RNase | Sample:lysis buffer volume ratio is too large, tissue is especially rich in RNases. | Redo with fresh sample, and add more lysis buffer, or decrease sample size, until you find a ratio that shows that all RNase has been deactivated. More lysis buffer is always better. |
| 10 | Nuclei are clumping when they are lysed, | Some cells are more delicate and clumping indicates that DNA is leaking out of the nuclei. | Try a shorter lysis, less igepal in the lysis buffer, or more BSA in the lysis buffer. Make sure that you included MgCl2 in the lysis buffer |
| 18 | Nuclei looked fine after lysis, but clumped after fixation. | Some cells are more delicate and clumping indicates that DNA is leaking out of the nuclei. | If sonicating the nuclei doesn't break up the clumps, try a shorter lysis, less igepal in the lysis buffer, and/or more BSA in the lysis buffer. Try fixing and rehydrating more gradually. |
| 44 | Losing too many nuclei when washing | | It's not unexpected to lose half the nuclei from the 2M that started in the RT plate. Recovery can be maximized by being mindful that every transfer of the nuclei will lose some to the walls of the tubes and pipettes. The supernatant doesn't always have to be completely removed for the washes if there is a chance to disturb the pellet. The biggest losses seem to happen when pooling wells, so at those steps make sure to gently pipet up/down a few times to dislodge settled nuclei before pulling them out of the well. |
| 67 | No smear of library on gel | Sample quality is the biggest factor, bad reagents. | If you've saved extra fixed nuclei, or extra final plates, you can retry from RT at step 20 or second strand synthesis at step 48 with fresh reagents. Optionally, take an aliquot of frozen nuclei and bulk RNA extract to make sure you are seeing RNA at all. Retry experiment with less tissue in the lysis. |

**Tiny Sci-RNA-Seq3**

This is the same sci protocol, but with the lysis and fixation scaled significantly down, with less transfers that will contribute to nuclei loss. We have used this method for E8.5 single embryos which are about 2-3mm in size, but it can more generally be used for instances where starting material is very limited. Ideally, samples (*e.g.* embryos or other) are isolated and frozen individually in 1.5 ml microfuge tubes with very little extra fluid with them.

You will need 100 µl lysis buffer per embryo. **This buffer is slightly different than the lysis buffer for older tissues.** Mix 5ml of the 10X-PBS-hypotonic stock solution, 75 µl of 2M MgCl2, and nuclease-free water to final volume of 50 ml to make the lysis base solution. *Right before lysis*, for every 1 ml of lysis buffer needed, add 5 µl 10% igepal, 40 µl BSA (20 mg/ml), 10 µl DEPC, then mix Vortex solution to disperse the DEPC throughout.

Add 100 µl of complete lysis buffer (with BSA/DEPC/igepal) to the tube with the frozen embryo, and make sure the embryo is actually in the buffer. <span style="color:red">**CAUTION**</span> you must work in the hood because of the DEPC in the lysis buffer. Let sit a couple of minutes on ice, then triturate the embryo gently with a pipette set to 50 µl with a yellow tip. You shouldn't see any chunks left. Lysis time is only about 5min. You can take 1 µl of this and mix with 9 µl of diluted yoyo dye to quickly make sure things are looking good at this point.

Mix fixative: 400 µl methanol + 10 µl DSP stock solution. Add 400 µl to the embryo, dripwise over a minute or two. Flick gently to mix and occasionally over 5-10 min on ice. You may see some clumping happening now.

Add 1 ml 0.3M SPBSTM, dripwise, slowly and mixing gently. Spin 500g, 3 min, 4°C. You should see a very tiny pellet. Remove all but about 50 µl of the supernatant, without disturbing the pellet. Resuspend in 500 µl 0.3M SPBSTM. If you got clumps from fixing, sonicate the tube for 12s on low. Spin again 500g 3min 4°C. Carefully remove supernatant and resuspend in SPBSTM so that the volume is 42.5 µl. Add 2.5 µl dNTPs and put 5 µl into each well of 1 column on a plate. Fill up the plate with more embryos or some other nuclei that you have a lot of, and continue with the sci protocol at the reverse transcription step 24 as normal.

**Application of Optimized sciRNA-seq3 Protocol to E16.5 Mouse Embryo**

To showcase the optimized sci-RNA-seq3 protocol, we describe here its application to a whole mouse embryo from embryonic day 16.5 (E16.5) of development. We set up C57BL/6J timed mouse breeding pairs; evidence of vaginal plug is counted as embryonic day 0.5. 16 days later, the dam was sacrificed. E16.5 embryos were then dissected and briefly rinsed in 1xPBS; the excess liquid was wicked away by Kimwipe before immediately flash freezing in LN2. Mice had *ad libitum* access to standard chow and water and were housed in the University of Washington Animal Research and Care Facility on a standard 12 hr light cycle. All procedures were approved by the UW IACUC (protocol number PROTO201800017). E16.5 embryos were stored in a foil pouch in LN2 storage tank until ready for dissociation.

Nuclei were isolated from a single E16.5 mouse with a hypotonic, phosphate-based, lysis buffer containing sucrose and DEPC, and fixed with a combination of DSP and methanol. Nuclei were washed and resuspended in SPBSTM buffer. ~4 million of these nuclei were processed with 2 plates of RT indexes, 2 plates of ligation indexes, and 3.5 plates of PCR indexes, as per the protocol described above. About 2000 nuclei per well were distributed into the final PCR plates. All 3.5 plates of PCR reactions were pooled and sequenced on an Illumina NovaSeq 6000 with dual index reads using the S4-200 kit and standard primers (Read1 34 cycles, Index1 10 cycles, Index2 10 cycles, Read2 100 cycles), resulting in about 7.15 billion reads in total, or 4.34 billion after removal of PCR duplicates.

After processing the sequencing data using the original sci-RNA-seq3 pipeline[1], we obtained profiles for 771,329 nuclei with UMI count per cell > 200. Even though these new data are still sequenced to a lower duplication rate relative to Cao *et al* (2019)[1] (39% and 46%, respectively), the optimized sci-RNA-seq3 method has markedly improved data quality, with ~4-fold higher UMIs and ~3-fold higher gene detection per nucleus (median UMI count 2,530; median genes detected 1,446; **Figure 8a**). We further filtered out cells which were detected as potential doublets, and then we set upper and lower thresholds of UMI counts used for quality filtering, which correspond to the mean +/- 2 standard deviations of log2-scaled values (except for the lower bound of 800, which was manually assigned). The resulting high-quality dataset included 381,888 cells that were further analyzed.

Read alignment and gene count matrix generation was performed using the pipeline that we developed for sci-RNA-seq3[1] with minor modifications: base calls were converted to fastq format using Illumina's bcl2fastq/v2.20 and demultiplexed based on PCR i5 and i7 barcodes using maximum likelihood demultiplexing package deML[19] with default settings. Downstream sequence processing and single cell digital expression matrix generation were similar to sci-RNA-seq[2] except that RT index was combined with hairpin ligation adaptor index, and thus the mapped reads were split into constituent cellular indices by demultiplexing reads using both the RT index and ligation index (Levenshtein edit distance (ED) < 2, including insertions and deletions). Briefly, demultiplexed reads were filtered based on RT index and ligation index (ED < 2, including insertions and deletions) and adaptor-clipped using trim_galore/v0.6.5 with default settings. Trimmed reads were mapped to the mouse reference genome (mm10) for mouse embryo nuclei, using STAR/v2.6.1d[20] with default settings and gene annotations (GENCODE VM12 for mouse). Uniquely mapping reads were extracted, and duplicates were removed using the unique molecular identifier (UMI) sequence (ED < 2, including insertions and deletions), reverse transcription (RT) index, hairpin ligation adaptor index and read 2 end-coordinate (*i.e.* reads with identical UMI sequence (less than edit distance of 2), RT index, ligation adaptor index and tagmentation site were considered duplicates). Finally, mapped reads were split into constituent cellular indices by further demultiplexing reads using the RT index and ligation hairpin index (ED < 2, including insertions and deletions). To generate digital expression matrices, we calculated the number of strand-specific UMIs for each cell mapping to the exonic and intronic regions of each gene with python/v2.7.13 HTseq package[21]. For multi-mapped reads, reads were assigned to the closest gene, except in cases where another intersected gene fell within 100 bp to the end of the closest gene, in which case the read was discarded. For most analyses, we included both expected-strand intronic and exonic UMIs in per-gene single-cell expression matrices.

After the single cell gene count matrix was generated, cells with low quality (UMI < 200 or detected gene > 100 or unmatched_rate ≥ 0.4) were filtered out and 771,329 cells were left. For the detection of potential doublet cells, we first split the dataset into subsets for each individual, and then applied the scrublet/v0.1 pipeline[22] to each subset with parameters (min_count = 3, min_cells = 3, vscore_percentile = 85, n_pc = 30, expected_doublet_rate = 0.06, sim_doublet_ratio = 2, n_neighbors = 30, scaling_method = 'log') for doublet score calculation. Cells with doublet score over 0.2 were annotated as detected doublets. We detected 2% potential doublet cells in the whole data set.

For detection of doublet-derived subclusters for cells, we used an iterative clustering strategy based on Scanpy/v.1.6.0[23]. Briefly, gene count mapping to sex chromosomes were removed before clustering and dimensionality reduction, and then genes with no count were filtered out and each cell was normalized by the total UMI count per cell. The top 1,000 genes with the highest variance were selected and the digital gene expression matrix was renormalized after gene filtering. The data was log transformed after adding a pseudocount, and scaled to unit variance and zero mean. The dimensionality of the data was reduced by PCA (30 components) first and then with UMAP, followed by Louvain clustering performed on the 30 principal components with default parameters. For Louvain clustering, we first fitted the top 30 PCs to compute a neighborhood graph of observations with local neighborhood number of 50 by scanpy.pp.neighbors. We then cluster the cells into sub-groups using the Louvain algorithm implemented as scanpy.tl.louvain function. For UMAP visualization, we directly fit the PCA matrix into scanpy.tl.umap function with min_distance of 0.1. For subcluster identification, we selected cells in each major cell type and applied PCA, UMAP, Louvain clustering similarly to the major cluster analysis. Subclusters with a detected doublet ratio (by Scrublet) over 15% were annotated as doublet-derived subclusters.

For data visualization, cells labeled as doublets (by Scrublet) or from doublet-derived subclusters were filtered out. For each cell, we only retain protein-coding genes, lincRNA genes and pseudogenes. Genes expressed in less than 10 cells and cells in which fewer than 100 genes were detected were further filtered out. The downstream dimension reduction and clustering analysis were done with Monocle/3-alpha. The dimensionality of the data was reduced by PCA (50 components) first on the top 5,000 most highly dispersed genes and then with UMAP (max_components = 2, n_neighbors = 50, min_dist = 0.1, metric = 'cosine'). Cell clusters were identified using the Louvain algorithm implemented in Monocle/3 (res = 1e-06). We generally find that the above Scrublet and iterative clustering based approach is limited in marking cell doublets between abundant cell clusters and rare cell clusters (e.g. less than 1% of total cell population). To further remove such doublet cells, we took the cell clusters identified by Monocle/3, downsampled each cell cluster to 2,500 cells, and computed differentially expressed genes across cell clusters with the top_markers function of Monocle/3 (reference_cells=1000). We then selected a gene set combining the top ten gene markers for each cell cluster (filtering out genes with fraction_expressing < 0.1 and then ordering by pseudo_R2). Cells from each main cell cluster were selected for dimension reduction by PCA (10 components) first on the selected gene set of top cluster specific gene markers, and then by UMAP (max_components = 2, n_neighbors = 50, min_dist = 0.1, metric = 'cosine'), followed by clustering identification using the Louvain algorithm implemented in Monocle/3 (res = 1e-04 for most clustering analysis). Subclusters showing low expression of target cell cluster-specific markers and enriched expression of non-target cell cluster-specific markers were annotated as doublets derived subclusters and filtered out in visualization and downstream analysis. We further filtered out the potential low-quality cells by investigating the numbers of UMIs and the proportion of reads mapping to the exonic regions per cell, resulting in a set of 381,888 cells that were used for performing dimension reduction.

We took the unique molecular identifiers (UMI) count matrix (feature ✕ nuclei) to perform conventional single-cell RNA-seq data processing using *Monocle*/v3: 1) normalizing the UMI counts by the estimated size factor per cell followed by log-transformation; 2) applying PCA and then using the top 50 PCs to perform UMAP dimension reduction (umap.n_neighbors = 50, umap.min_dist = 0.01, max_components = 2); 3) performing louvain clustering using cluster_cells function in *Monocle*/v3.

We next performed manual annotation of individual clusters based on marker gene expression. As shown in a 2D UMAP, cells were generally assigned with one of 20 major developmental trajectories (**Figure 8b**). As compared to trajectories described in Cao *et al.* (2019)[1], which were generated on mice ranging from E9.5 to E13.5, we identified several new cell types (*e.g.* adipocytes) and relatively dense substructures for some major trajectories. Focussing in further on white blood cells, these can be further separated into multiple sub-trajectories, including T cells, B cells, different types of macrophages, *etc.* (**Figure 8b**). Of note, we observe closely related but distinct populations of cells corresponding to border-associated macrophages (*Lyve1*+, *F13a1*+) and microglia (*Sall1*+, *Sall3*+), consistent with a previous study of the same stage of mouse development[24].

## Summary

In our hands, this simplified, optimized sci-RNA-seq3 protocol is faster, higher yield, more robust, more sensitive, and more cost effective than the original sci-RNA-seq3 protocol. The protocol is also adaptable to very small sample inputs. Additional modifications to streamline or even remove the tagmentation step may further improve the protocol, which can also potentially be combined with oligo-based hashing techniques[16,17]. Finally, we note that some of the optimizations reported here (*e.g.* fixation conditions, etc.) may be useful for improving the performance of other sci-* methods, as well as for other single cell profiling technologies.

## Acknowledgements

We thank Bridget Kulesakara for the sucrose buffer advice, Jase Gehring for fixative expertise and Riza Daza for helpful feedback. Diana O'Day, Mai Le, Roshella Gomes, Saskia Ilcisin, Dana Jackson for protocol testing, and the entire Shendure Lab for support and encouragement. This work was funded in part by NIH R01HG010632 to JS and CT, UM1HG011586 to JS and BB, R35GM137916 to BB and T32HG000035 to EN. JS is an Investigator of the Howard Hughes Medical Institute.

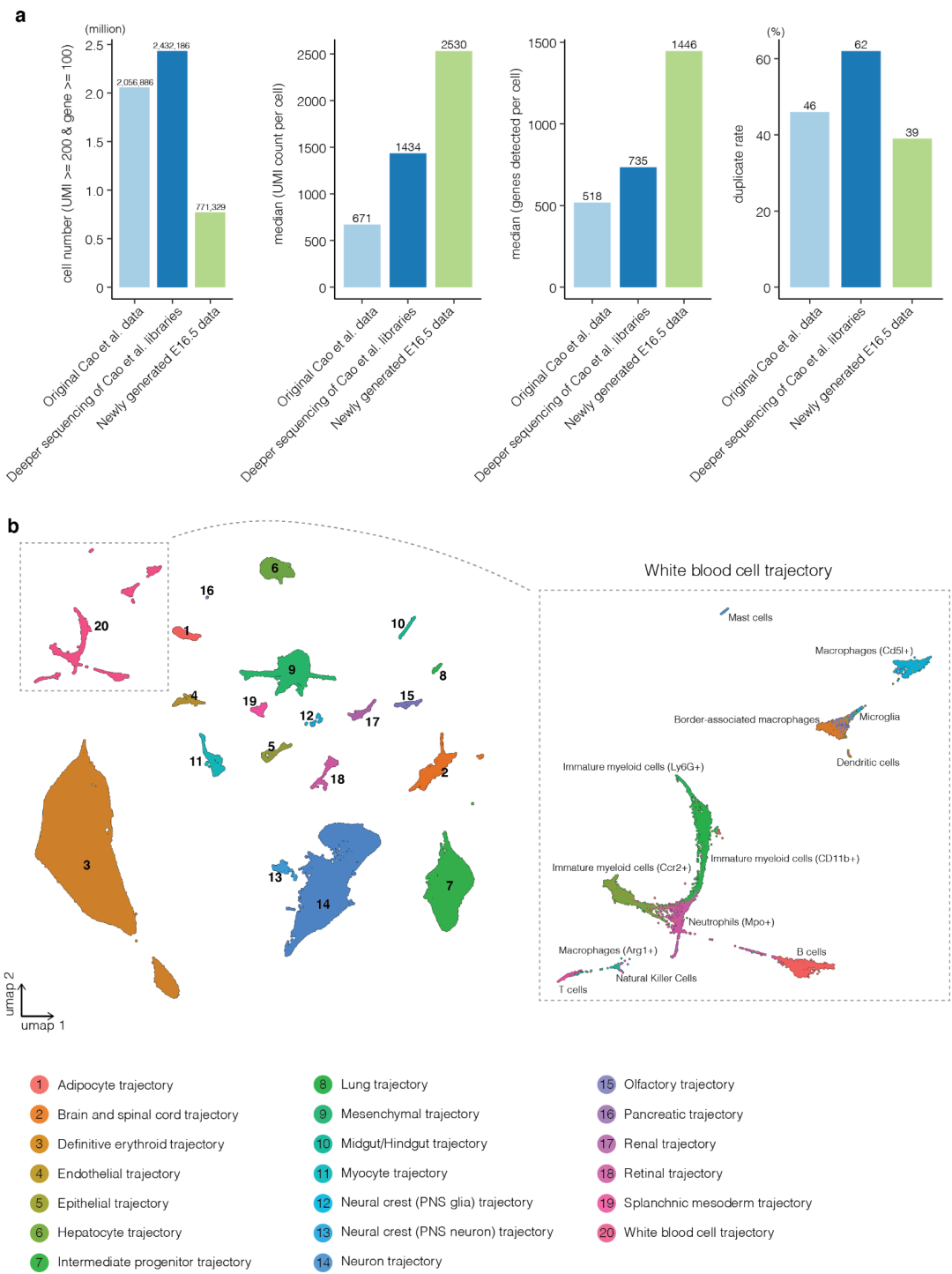

**Figure 9. High-quality data of E16.5 mouse embryo generated by application of the optimized sci-RNA-seq3 protocol. a,** The cell number, median UMI count per cell, median genes detected per cell, and duplicate rate, are shown for a previously published dataset on E9.5 - E13.5 embryos (light blue bars)[1], deeper sequencing and reanalysis of those same sequencing libraries (dark blue bars) or data newly generated on E16.5 embryo using the optimized sci-RNA-seq3 protocol (green bars). **b,** 2D UMAP visualization of the new E16.5 dataset. All nuclei colored by each of the 20 cell trajectories are shown on the left. Subview of global 2D UMAP visualization highlighting subpopulations of the white blood cells trajectory is shown on the right.

**References**

1.  Cao, J. *et al.* The single-cell transcriptional landscape of mammalian organogenesis. *Nature* **566**, 496–502 (2019).

2.  Cao, J. *et al.* Comprehensive single-cell transcriptional profiling of a multicellular organism. *Science* **357**, 661–667 (2017).

3.  Domcke, S. *et al.* A human cell atlas of fetal chromatin accessibility. *Science* **370**, (2020).

4.  Cusanovich, D. A. *et al.* Multiplex single cell profiling of chromatin accessibility by combinatorial cellular indexing. *Science* **348**, 910–914 (2015).

5.  Rosenberg, A. B. *et al.* Single-cell profiling of the developing mouse brain and spinal cord with split-pool barcoding. *Science* **360**, 176–182 (2018).

6.  Datlinger, P. *et al.* Ultra-high throughput single-cell RNA sequencing by combinatorial fluidic indexing. doi:10.1101/2019.12.17.879304.

7.  Lareau, C. A. *et al.* Droplet-based combinatorial indexing for massive-scale single-cell chromatin accessibility. *Nat. Biotechnol.* **37**, 916–924 (2019).

8.  Vitak, S. A. *et al.* Sequencing thousands of single-cell genomes with combinatorial indexing. *Nat. Methods* **14**, 302–308 (2017).

9.  Yin, Y. *et al.* High-Throughput Single-Cell Sequencing with Linear Amplification. *Mol. Cell* **76**, 676–690.e10 (2019).

10. Ramani, V. *et al.* Massively multiplex single-cell Hi-C. *Nat. Methods* **14**, 263–266 (2017).

11. Mulqueen, R. M. *et al.* Highly scalable generation of DNA methylation profiles in single cells. *Nat. Biotechnol.* **36**, 428–431 (2018).

12. Cao, J. *et al.* Joint profiling of chromatin accessibility and gene expression in thousands of single cells. *Science* **361**, 1380–1385 (2018).

13. Cao, J., Zhou, W., Steemers, F., Trapnell, C. & Shendure, J. Sci-fate characterizes the dynamics of gene expression in single cells. *Nat. Biotechnol.* **38**, 980–988 (2020).

14. Hwang, B. *et al.* SCITO-seq: single-cell combinatorial indexed cytometry sequencing. *Nat. Methods* **18**, 903–911 (2021).

15. Wang, Q. *et al.* CoBATCH for High-Throughput Single-Cell Epigenomic Profiling. *Mol. Cell* **76**, 206–216.e7 (2019).
16. Srivatsan, S. R. *et al.* Massively multiplex chemical transcriptomics at single-cell resolution. *Science* **367**, 45–51 (2020).
17. Srivatsan, S. R. *et al.* Embryo-scale, single-cell spatial transcriptomics. *Science* **373**, 111–117 (2021).
18. Cao, J. *et al.* A human cell atlas of fetal gene expression. *Science* **370**, (2020).
19. Renaud, G., Stenzel, U., Maricic, T., Wiebe, V. & Kelso, J. deML: robust demultiplexing of Illumina sequences using a likelihood-based approach. *Bioinformatics* **31**, 770–772 (2015).
20. Dobin, A. *et al.* STAR: ultrafast universal RNA-seq aligner. *Bioinformatics* **29**, 15–21 (2013).
21. Anders, S., Pyl, P. T. & Huber, W. HTSeq--a Python framework to work with high-throughput sequencing data. *Bioinformatics* **31**, 166–169 (2015).
22. Wolock, S. L., Lopez, R. & Klein, A. M. Scrublet: Computational Identification of Cell Doublets in Single-Cell Transcriptomic Data. *Cell Syst* **8**, 281–291.e9 (2019).
23. Wolf, F. A., Angerer, P. & Theis, F. J. SCANPY: large-scale single-cell gene expression data analysis. *Genome Biol.* **19**, 15 (2018).
24. Utz, S. G. *et al.* Early Fate Defines Microglia and Non-parenchymal Brain Macrophage Development. *Cell* **181**, 557–573.e18 (2020).